\documentclass[%
reprint,
twocolumn,
superscriptaddress,
showpacs,preprintnumbers,
nofootinbib,
aps,
prl,
]{revtex4-1}

\usepackage{tikz}
\usepackage{amsmath}
\usepackage{amssymb}
\usepackage{mathtools}
\usepackage{hyperref}
\usepackage{graphicx}
\usepackage{ulem}
\usepackage{comment}

\graphicspath{
{./}
{./figures/}
{./figures/sketch/}
{./figures/isolated/}
}

\def\ga{\gamma}

\def\th{\theta}

\def\ka{\kappa}
\def\la{\lambda}

\def\si{\sigma}
\def\ta{\tau}

\newcommand{\ben}{\begin{equation}}
\newcommand{\een}{\end{equation}}
\newcommand{\bea}{\begin{eqnarray}}
\newcommand{\eea}{\end{eqnarray}}
\newcommand{\ba}{\begin{array}}
\newcommand{\ea}{\end{array}}
\newcommand{\bit}{\begin{itemize}}
\newcommand{\eit}{\end{itemize}}

\newcommand{\half}{\frac12}

\newcommand{\mMon}{M_\text{m}}

\newcommand{ \Tr}{\mathrm{Tr} \,} 

\begin{document}

\newcommand{\Sussex}{\affiliation{
Department of Physics and Astronomy,
University of Sussex, Falmer, Brighton BN1 9QH,
U.K.}}

\newcommand{\HIPetc}{\affiliation{
Department of Physics and Helsinki Institute of Physics,
PL 64,
FI-00014 University of Helsinki,
Finland
}}

\newcommand{\Stavanger}{\affiliation{
Institute of Mathematics and Natural Sciences,
University of Stavanger,
4036 Stavanger,
Norway
}}

\title{New solutions for non-Abelian cosmic strings}

\author{Mark Hindmarsh}
\email{m.b.hindmarsh@sussex.ac.uk}
\Sussex
\HIPetc
\author{Kari Rummukainen}
\email{kari.rummukainen@helsinki.fi}
\HIPetc
\author{David J. Weir}
\email{david.weir@uis.no}
\Stavanger

\date{\today}

\begin{abstract}
We study the properties of classical vortex solutions in a non-Abelian
gauge theory. A system of two adjoint Higgs fields breaks the
$\mathrm{SU}(2)$ gauge symmetry to $Z_2$, producing 't Hooft-Polyakov
monopoles trapped on cosmic strings, termed beads; there are two
charges of monopole and two degenerate string solutions.  The strings
break an accidental discrete $Z_2$ symmetry of the theory, explaining
the degeneracy of the ground state.  Further symmetries of the model,
not previously appreciated, emerge when the masses of the two adjoint
Higgs fields are degenerate. The breaking of the enlarged discrete
symmetry gives rise to additional string solutions and splits the
monopoles into four types of `semipole': kink solutions that
interpolate between the string solutions, classified by a complex
gauge invariant magnetic flux and a $Z_4$ charge.  At special values
of the Higgs self-couplings, the accidental symmetry broken by the
string is continuous, giving rise to supercurrents on the strings.
The $\mathrm{SU}(2)$ theory can be embedded in a wide class of Grand
Unified Theories, including $\mathrm{SO}(10)$. We argue that semipoles
and supercurrents are generic on GUT strings.
\end{abstract}

\preprint{HIP-2016-21/TH}
\maketitle


Cosmic strings \cite{Kibble:1976sj} are line-like concentrations of
energy and tension which may have been created in the early universe
(for reviews see
\cite{Hindmarsh:1994re,Vilenkin:2000jqa,Copeland:2011dx,Hindmarsh:2011qj}). They
may be either classical vortex solutions in a field theory with
spontaneously broken symmetries \cite{Kibble:1976sj}, or fundamental
objects in a string theory
\cite{Witten:1985fp,Sarangi:2002yt,Copeland:2003bj}.  The form of the
Standard Model as a theory of gauge (local) symmetries spontaneously
broken at a scale of O(100) GeV, and its inability to account for dark
matter and the baryon asymmetry of the universe, motivates the study
of extensions to the Standard Model with further spontaneously broken
gauge symmetries.

The simplest such extension, a theory with an extra gauged
$\mathrm{U}(1)$ symmetry, has cosmic string solutions in the form of
Nielsen-Olesen vortices \cite{Nielsen:1973cs} in the Abelian Higgs
model. These Abelian Higgs strings have been extensively studied
numerically in order to establish the dynamics of their formation and
evolution in the early universe, and the ensuing observational
predictions
\cite{Laguna:1989hn,Vincent:1997cx,Moore:2001px,Bevis:2006mj,Bevis:2010gj}.
Theories with two broken $\mathrm{U}(1)$ gauge symmetries, in which strings can
form bound states and junctions, have been studied numerically as a
model for cosmic fundamental strings
\cite{Urrestilla:2007yw,Lizarraga:2016hpd}.

Other spontaneously broken symmetries produce strings.  Indeed, if a
compact non-abelian gauge group $G$ is broken to a subgroup $H,$
strings are guaranteed if the manifold $G/H$ is non-simply connected
\cite{Kibble:1976sj}. If $G$ is itself simply connected, it can be
shown that $G/H$ is non-simply connected if and only if $H$ is
disconnected.  The simplest class of examples is the symmetry-breaking
$\mathrm{SU}(2) \to Z_N$, and string network simulations have been performed in
theories with broken global symmetries and $N = 3$
\cite{Hindmarsh:2006qn}.

In the absence of global symmetries and accidental degeneracies,
$G/H$ is isomorphic to the vacuum manifold of the theory. If $G$ is
embedded in a larger global symmetry group, the strings may be
``semi-local'' \cite{Vachaspati:1991dz} and unstable if the scalar
self-coupling is large enough \cite{Hindmarsh:1992yy}.

In this paper we study strings in a non-abelian gauge theory, with
symmetry-breaking $\mathrm{SU}(2) \to Z_2$.  A string solution in this
theory was also discovered by Nielsen and Olesen
\cite{Nielsen:1973cs}, and its detailed structure later elucidated in
Refs.~\cite{Hindmarsh:1985xc,deVega:1986eu,deVega:1986hm,Aryal:1987sn}.
A major interest of this model is that it can be embedded naturally in
Grand Unified Theories (GUTs) such as $\mathrm{SO}(10)$
\cite{Kibble:1982ae}. Furthermore, it has been argued that cosmic
strings are generic in such GUTs~\cite{Jeannerot:2003qv}.  It can also
be modified to allow two symmetry-breaking scales with an intermediate
unbroken $\mathrm{U}(1)$ symmetry, modelling a two-stage GUT
symmetry-breaking.  The first stage, $\mathrm{SU}(2) \to
\mathrm{U}(1)$, produces 't Hooft-Polyakov
monopoles~\cite{Hooft:1974qc,Polyakov:1974ek}, and the second confines
the monopole flux to two strings.  The resulting composite object is
called a bead \cite{Hindmarsh:1985xc}. A system of many monopoles
trapped on a string is known as a necklace
\cite{Berezinsky:1997td}. The evolution of a system of necklaces could
be quite different from ordinary cosmic strings, unless beads
annihilate efficiently~\cite{Siemens:2000ty,BlancoPillado:2007zr},
with the potential for important cosmic ray and $\ga$-ray signals.
Monopoles on strings have recently been reviewed in
\cite{Kibble:2015twa}.

In this paper we elucidate the importance of global symmetries in the
classification of the beads, and discover new solutions which we call
semipoles. In the case with two-stage symmetry breaking, there is a
$Z_2$ symmetry spontaneously broken by the string solutions, and beads
can be viewed as the resulting kinks.  This was not fully appreciated
before.  The new solutions appear in the case where the
$\mathrm{SU}(2)$ and $\mathrm{U}(1)$ symmetry-breaking scales are
degenerate, giving rise to an enlarged discrete global symmetry. The
monopoles and antimonopoles each split into two semipoles,
interpolating between four degenerate string solutions, which can be
thought of as sources of a complex gauge invariant magnetic flux.  The
discrete symmetry can be further promoted to a global $\mathrm{O}(2)$
symmetry at a special value of the cross-coupling between the scalar
fields. This symmetry is spontaneously broken by the string solution
but not the vacuum.  Hence the strings carry persistent global
currents, rather like a trapped superfluid
\cite{Witten:1984eb,Alford:1990mk,Hindmarsh:1991ax}.

We study the $\mathrm{SU}(2)$ Georgi-Glashow model with two Higgs
fields. This model has the Lagrangian
\begin{equation}
\mathcal{L} = -\frac{1}{4}  F_{\mu\nu}^a F^{\mu\nu a} + \sum_n \mathrm{Tr}\, [D_\mu,\Phi_n][D^\mu,\Phi_n] - V(\Phi_1, \Phi_2)
\label{e:SU2HLag}
\end{equation}
where $D_\mu = \partial_\mu + i g A_\mu$ is the covariant derivative,
with $F_{\mu\nu} = F^a_{\mu\nu} \tau^a$, $A_\mu = A_\mu^a \tau^a$,
$\tau^a = \sigma^a/2$ where $\sigma^a$ is a Pauli matrix. The Higgs
fields $\Phi_n$, $n=1,2$, are in the adjoint representation, $\Phi_n =
\phi_n^a \tau^a$.

The potential is
\begin{multline}
\label{eq:potential}
V(\Phi_1, \Phi_2) = m_1^2 \mathrm{Tr} \, \Phi_1^2 + \lambda (\mathrm{Tr} \,
\Phi_1^2)^2 + m_2^2 \mathrm{Tr} \, \Phi_2^2 + \lambda (\mathrm{Tr} \,
\Phi_2^2)^2 \\ 
+ \kappa (\mathrm{Tr} \, \Phi_1 \Phi_2)^2,
\end{multline}
with $\la$ and $\ka$ positive.
This form is motivated by the the SO(10) GUT embedding, and 
the wish to investigate separate scales for the strings and monopoles without unnecessarily complicating the parameter space.
A fully general potential does not change the following discussion.
The directions of the vacuum expectation values
(vevs) are perpendicular, because of the $(\mathrm{Tr} \, \Phi_1
\Phi_2)^2$ term in the potential.  The system therefore undergoes two
symmetry-breaking phase transitions, $\mathrm{SU}(2) \to \mathrm{U}(1)
\to Z_2$, as the parameters $m_{1,2}^2$ are given negative values
sequentially.  The vevs of the two adjoint scalar fields are given by
$\mathrm{Tr} \, \Phi_{1,2}^2 = \left| m_{1,2}^2 \right|/2 \lambda$, or
$v_{1,2}^2 = \left| m_{1,2}^2 \right|/ \lambda$. The scalar masses are
then $\sqrt{2} m_{1,2}$.  Without loss of generality, we can label the
scalar fields such that $\Phi_1$ has the larger vacuum expectation
value, and is responsible for the first of the symmetry-breakings by
taking the value $\Phi_1 = v_1\ta^3$.

The vev of $\Phi_2$, which can be taken as $\Phi_2 = v_2 \tau^1$,
breaks the remaining gauge symmetry $\mathrm{U}(1) \to Z_2$, after
which all the gauge bosons are massive.  The remaining $Z_2$ symmetry
is comprised of the discrete gauge transformations $U = \pm1$.  After both symmetry-breakings
have taken place, the gauge bosons have masses (in decreasing order)
$g \sqrt{v_1^2 + v_2^2}$, $g v_1$ and $g v_2$.
  
The symmetry-breaking $\mathrm{SU}(2) \to Z_2$ produces topological
defects in the form of vortices.  If one considers the two
symmetry-breakings individually, the first produces 't Hooft-Polyakov
monopoles, while the second produces strings which carry the flux
associated with the monopoles.  From numerical
measurements~\cite{Forgacs:2005vx}, the classical monopole mass is
\begin{equation}
\label{eq:monopolemass}
\mMon = \frac{4\pi v_1}{g}
f_\mathrm{m}\left(\frac{2\lambda}{g^2}\right); \qquad f_\mathrm{m}(1)
\approx 1.238.
\end{equation}

The solution for a string oriented along the $z$ axis can be written
in cylindrical polar coordinates $\rho,$ $\th$ as
\cite{Hindmarsh:1985xc,Aryal:1987sn}
\begin{align}
\label{e:StrSolPhi1}
\Phi_1 &\to v_1 k(\rho)  \tau^3\\
\label{e:StrSolPhi2}
\Phi_2 &\to v_2  h(\rho) (\tau^1\cos\th + \tau^2\sin\th),\\
\label{e:StrSolA}
A_i &\to \hat\th_i \frac{a(\rho)}{g\rho} \ta^3 .
\end{align}
The functions $a$ and $h$ must vanish at the origin, and tend to unity
at infinity.  The function $k$ tends to $\pm 1$, and does not in
general vanish at the origin.

On substitution of this ansatz into the Lagrangian, one can see that
the $\Phi_1$ and $\Phi_2$ fields decouple completely.  In this case,
$k(\rho) = \pm 1$ everywhere, and the functions $a(\rho)$ and
$h(\rho)$ become those for a Nielsen-Olesen vortex.  The string
tension is
\begin{equation}
\label{eq:stringtension}
\mu = {\pi}v_2^2 f_\mathrm{s}\left(\frac{2\lambda}{g^2}\right), 
\end{equation}
where $f_\mathrm{s}(1) = 1$.

\begin{figure*}
\begin{center}

\includegraphics[width=\textwidth]{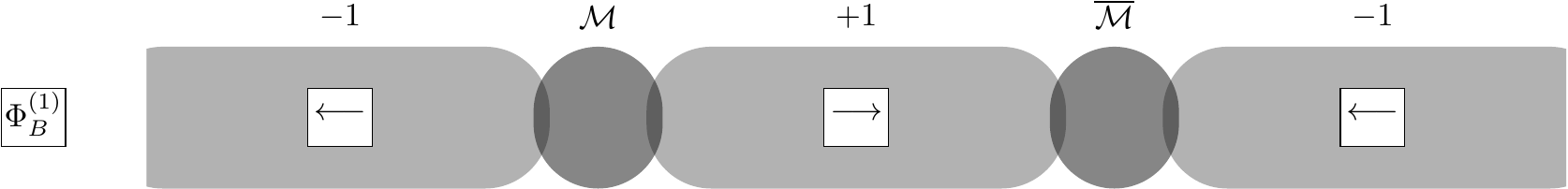} \\
\smallskip
\includegraphics[width=\textwidth]{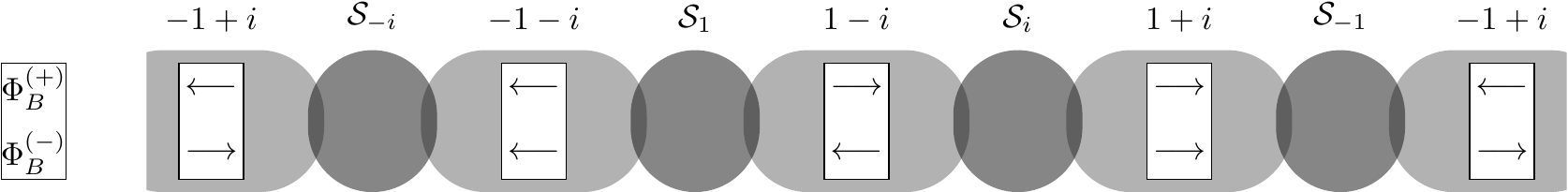}

\end{center}
\caption{\label{fig:sketch} Sketch of monopole and semipole solutions
  on a straight string. At top is the situation when $m_1^2 > m_2^2$,
  showing segments of string (light elongated shapes) with monopole
  `beads' (dark circular shapes). The direction of the flux
  $\Phi_B^{(1)}$ (\ref{e:ComFluB}) along the string segments is
  indicated by the direction of the arrows.  At bottom, the case when
  $m_1^2=m_2^2$ and $\kappa > 2\lambda$ is shown, with the dark
  circular shapes representing semipoles.  Here, the arrows indicate
  the real and imaginary parts of the gauge invariant complex flux,
  Eq.~(\ref{e:ComFluA}). There are four possible string solutions
  labelled by $w$, with semipoles $\mathcal{S}$ interpolating between
  them where the subscripts give 
  their complex flux. In the case where
  $\kappa < 2\lambda$, the appearance is similar but with
  $\Phi_B^{(1,2)}$ replacing $\Phi_B^{(\pm)}$. Note that the
  arrangement of semipoles shown is unable to annihilate; other
  orderings would allow annihilation.}
\end{figure*}

\begin{figure}
\begin{center}
\fbox{\includegraphics[trim={0 3cm 0 3cm},clip,width=0.47\textwidth]{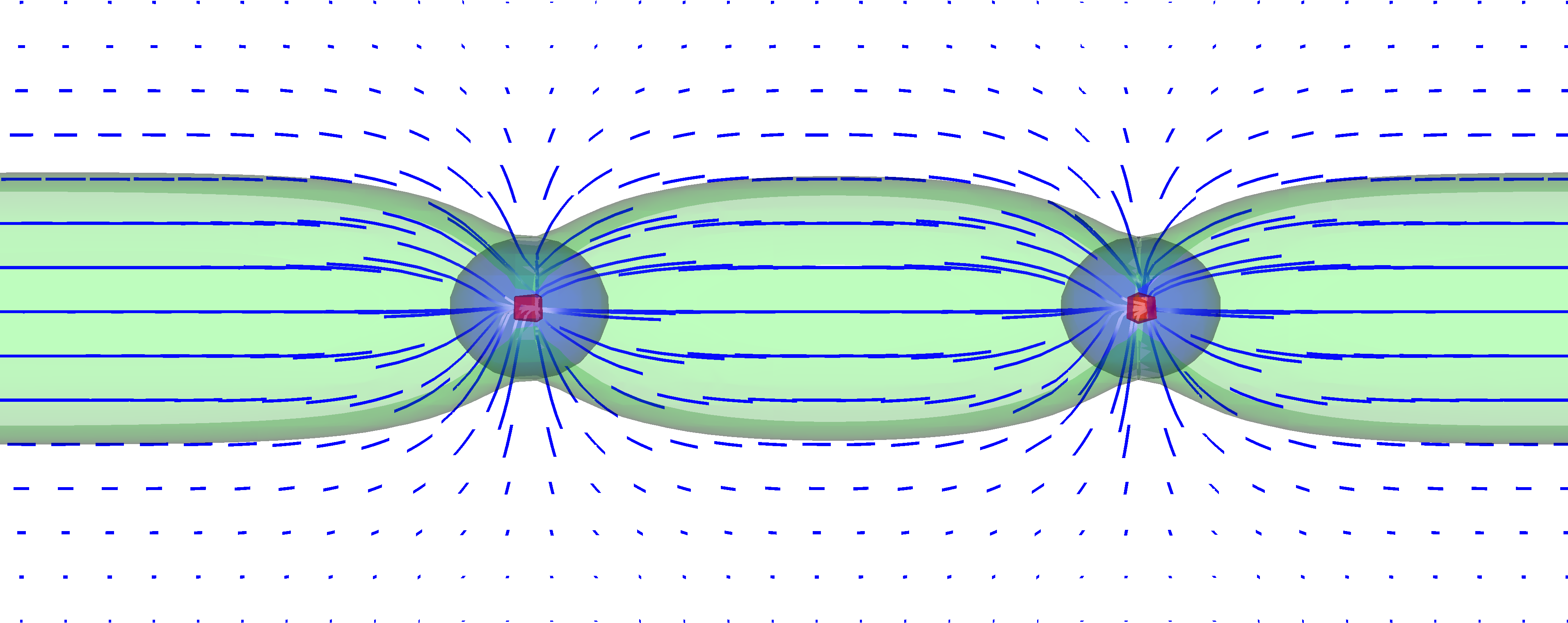}} \\
\medskip

\fbox{\includegraphics[trim={0 3cm 0 3cm},clip,width=0.47\textwidth]{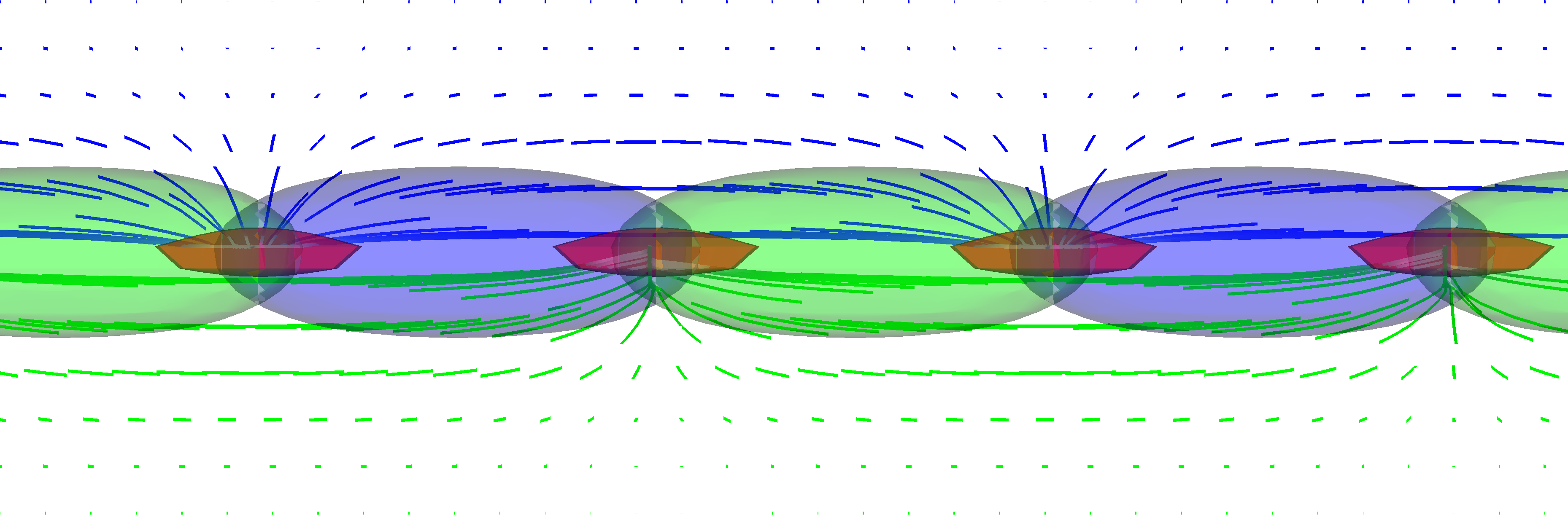}} \\
\medskip

\fbox{\includegraphics[trim={0 3cm 0 3cm},clip,width=0.47\textwidth]{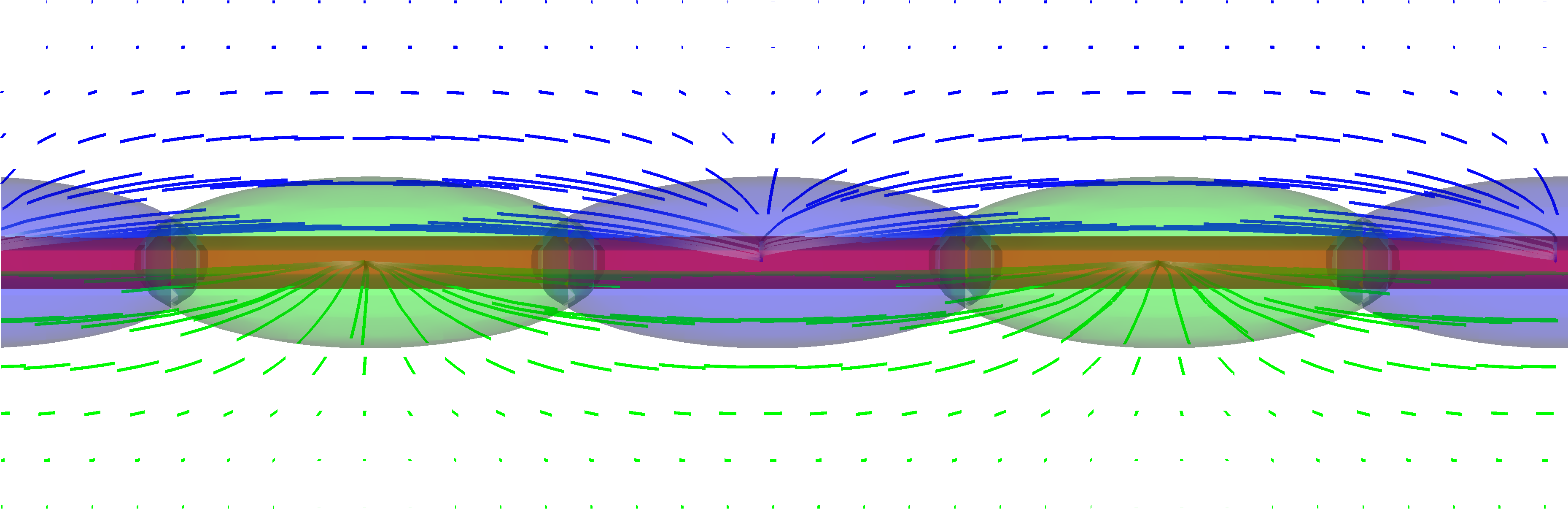}} \\
\medskip

\fbox{\includegraphics[trim={0 3cm 0 3cm},clip,width=0.47\textwidth]{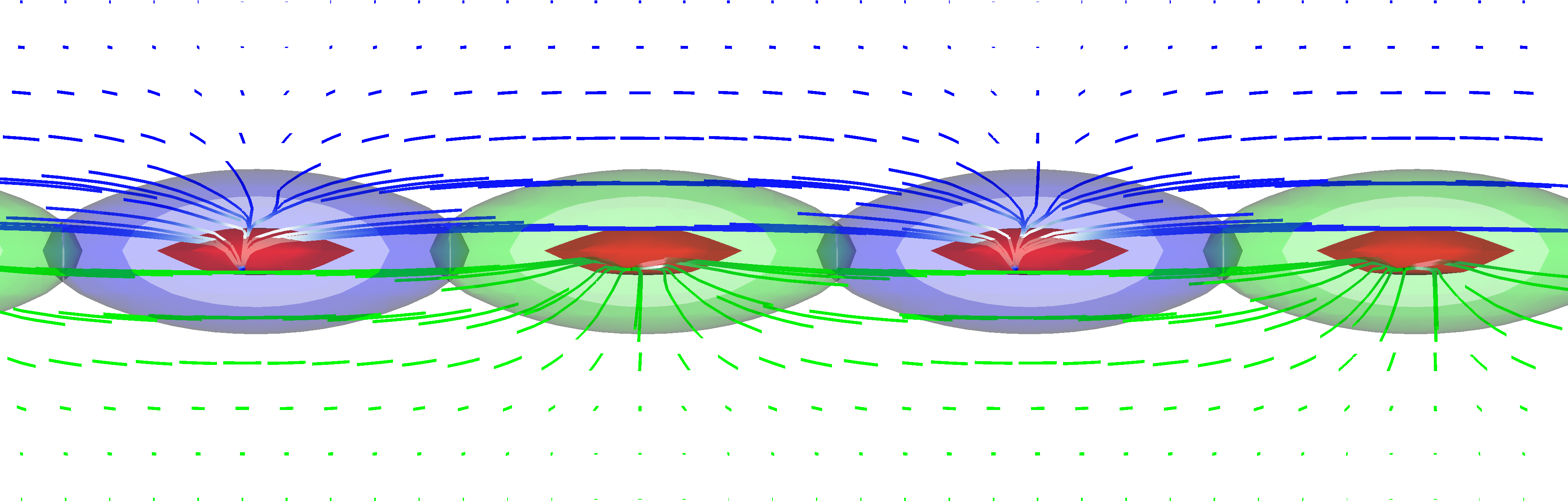}} \\
\end{center}

\caption{\label{fig:isolated} String segments from numerical
  solutions, showing isosurfaces of $\Tr \, \Phi_{1,2}^2 =
  \frac{5}{16}v^2_{1,2}$ in blue and green respectively, and 90\%
  isosurfaces of energy density in red. At the top, $(a m_1)^2 = 0.1$,
  $(a m_2)^2=0.04$ and $\kappa=1$. The $B^{(1)}$ field
  (\ref{magfield}) is shown as blue streamlines on a vertical slice
  through the string, their length proportional to the field intensity
  at a given point. The other three images have $(a m_1)^2 = (a m_2)^2
  = 0.1$. Second to top is $\kappa = 2$: Here, the $B^{(+)}$ and
  $B^{(-)}$ fields (\ref{magfieldPM}) are shown in blue and green,
  above and below the string (note that the streamlines are axially
  symmetric).  The lower two images are of numerical solutions with
  $\kappa=1$ and $\ka = 0.5$, again showing $B^{(1)}$ streamlines in
  blue above the string, and the $B^{(2)}$ streamlines in green
  below.}
\end{figure}

Note that the strings carry half the flux of a monopole, and monopoles
can therefore be thought of as threaded like a bead on a string
\cite{Hindmarsh:1985xc}.

The theory of Eq.~(\ref{e:SU2HLag}) has a number of global symmetries,
which are important when enumerating the multiplicity of static
solutions.  Normally, we study the symmetries of the vacuum
configuration, which has constant scalar field and zero gauge field
strength, and we ignore transformations on the gauge field.  However,
we are also interested in the multiplicity of the vortex solutions,
and so must be careful to include transformations on both scalar and
gauge fields.

In the general case, the theory has a discrete global $Z_2\times Z_2$
symmetry $\Phi_{1} \to \pm \Phi_{1}$ and $\Phi_{2} \to \pm \Phi_{2}$,
with the gauge field $A_\mu$ left alone.  We note that it is always
possible to find a gauge transformation which flips the signs of the
scalar fields simultaneously: $\Phi_n \to -\Phi_n$.  However, this
gauge transformation will also change the gauge field.  The difference
between the global reflection and gauge transformations can be made
explicit by considering the effective magnetic field operators
\begin{equation}
B^{(1)}_i = {\rm Tr} \, B_i \hat\Phi_1 ,
~~
B^{(2)}_i = {\rm Tr} \, B_i \hat\Phi_2 ,
\label{magfield}
\end{equation}
where $\hat\Phi_{1,2} = \Phi_{1,2} / \sqrt{{\rm Tr} \, (\Phi_{1,2}^2)}$
and $B_i = \half \epsilon_{ijk} F^{jk}$.   These are gauge
invariant, but obviously change sign under the reflection of the
scalar fields.  These operators are used to quantify the magnetic flux
in string solutions, 
along with a second set of gauge invariant magnetic  fields
\begin{equation}
B^{(\pm)}_i = {\rm Tr} \, B_i \hat\Phi_\pm ,
\label{magfieldPM}
\end{equation}
where 
\begin{equation}
\hat\Phi_\pm = \frac{1}{\sqrt{2}}\left(\hat\Phi_1 \pm \hat\Phi_2 \right).
\end{equation}

In the case of degenerate masses, $m_1^2 = m_2^2 = m^2$, the Lagrangian is
invariant under interchange of the fields, $\Phi_1 \leftrightarrow
\Phi_2$, and so the discrete symmetry is enhanced. It can be seen that
the sign change and interchange symmetries generate $D_4$, the group
of symmetries of a square.  Again, in a state with no gauge field
strength (or $A_\mu \perp \Phi_n$), a sign change on the scalar fields
is equivalent to a gauge transformation.

Given our specific form of potential~(\ref{eq:potential}), setting
$\kappa = 2\lambda$ means that the discrete symmetry is further
enhanced to a global $\mathrm{O}(2)$ symmetry.  This is most easily
expressed in terms of a complexified $\mathrm{SU}(2)$ triplet $\Phi =
(\Phi_1 + i \Phi_2)$ for which
\begin{multline}
V
= m^2 \, \mathrm{Tr} \, \Phi \Phi^\dag 
+  \frac{2\la - \ka}{16} \left(  \left( \mathrm{Tr} \, \Phi^2 \right)^2 + \left( \mathrm{Tr} \, \Phi^{\dag\,2} \right)^2 \right)
\\ 
+  \frac{\la}{2}\left(\mathrm{Tr} \, \Phi \Phi^\dag \right)^2 
+ \frac{2\lambda+\ka}{8} \left(\mathrm{Tr} \, \Phi^2\right)\left(\mathrm{Tr} \, \Phi^{\dag\,2}\right).
\end{multline}
When $\kappa = 2\lambda$ there are symmetry transformations 
\begin{equation}
\label{e:U1Sym}
\Phi \to e^{i\alpha}\Phi \quad \text{and} \quad \Phi \to \Phi^\dag
\end{equation}
which are seen to generate the group $\mathrm{O}(2)$. The sign change
$\Phi \to -\Phi$ is equivalent to a large gauge transformation on a
state with no gauge field strength.

The global symmetries are not broken in the vacuum, as their
transformations can be undone with a large gauge transformation.
For example, choosing the gauge in which the vacuum is
$(\Phi_1,\Phi_2) = (v_1\ta^3,v_2\ta^1)$, the effect of the change of
sign of $\Phi_1$ can be undone with a gauge transformation $U =
i\si^1$, while $U = i\si^3$ undoes the effect of a sign change on
$\Phi_2$.  In the degenerate case, the field interchange $\Phi_1
\leftrightarrow \Phi_2$ can be undone with a gauge transformation $U =
i(\si^1 + \si^3)/\sqrt{2}$.

However, some of the global symmetries are broken by the string
solutions. We recall there are three cases: general, degenerate
masses, and full global $\mathrm{O}(2)$ symmetry.  In the first
two cases there are two and four degenerate string solutions,
separated by finite energy barriers.  In the last case, there is a
one-parameter family of string solutions.

To see that the string solution breaks the discrete symmetry in the
general case, we use the magnetic field operator $B_i^{(1)}$, equation
(\ref{magfield}).  It is clearly odd under the $Z_2$ symmetry $\Phi_1
\to -\Phi_1$, and non-zero in the core of the string.  Strings
oriented in the $z$-direction can therefore be labelled by the sign of
the magnetic flux $\Phi_B^{(1)} = \pm 2\pi/g$, with
\begin{equation}
\Phi_B^{(1,2)} = \int d^2x \, B^{(1,2)}_i\hat{z}_i .
\end{equation}

In the mass degenerate case, we find three different types of string
solutions depending on relative size of $\kappa$ and $2\lambda$.  When
$\kappa > 2\lambda$, either $\Phi_1$ or $\Phi_2$ may vanish at the
centre, meaning that there are two string solutions which, in a
suitable gauge, may be written as
(\ref{e:StrSolPhi1}--\ref{e:StrSolA}) with $\Phi_1$ and $\Phi_2$
possibly interchanged.  The orientation of the magnetic field along
the string further divides the string solution to give four types.

In detail, we can use the magnetic field operators (\ref{magfieldPM})
to define a complex flux
\begin{equation}
\label{e:ComFluA}
\Phi _B = \Phi_B^{(+)} + i \Phi_B^{(-)} = w\frac{2\pi}{g}.
\end{equation}
If $\Phi_1$ vanishes at the centre of the string, $\Phi_B^{(-)} =
-\Phi_B^{(+)} = \pm2\pi/g$, while if $\Phi_2$ vanishes at the centre
of the string, $\Phi_B^{(-)} = \Phi_B^{(+)} =\pm2\pi/g$.  Hence $w =
(\pm1 \pm i)$.

We see that string solutions are in one-to-one correspondence with the
$Z_2 \times Z_2$ group generated by sign changes of the fields. They
leave unbroken a $Z_2$ subgroup of the original global symmetry $D_4$.

In the degenerate case with $\kappa < 2\lambda$, it is energetically
favourable for $\Phi_1$ and $\Phi_2$ to align or antialign in the core
of the string, with either $\Phi_B^{(+)}$ or $\Phi_B^{(-)}$ vanishing.
We can therefore label strings with the complex flux
\begin{equation}
\label{e:ComFluB}
\Phi'_B =  \Phi_B^{(1)} + i \Phi_B^{(2)} = w'\frac{2\pi}{g}.
\end{equation}
Again,  $w' = (\pm1 \pm i)$.

The string solutions are separated by finite-energy barriers, and so
solutions can be constructed which interpolate between the string
ground states as a function of $z$.  We display sketches of these
solutions in Fig.\ \ref{fig:sketch}.  In the general case, where there
are two string ground states, the kink or antikink solutions
interpolating between them are sources of flux which are the
difference between the fluxes of the adjacent strings, or
\begin{equation}
\Phi_B^{(1)} = \pm 4\pi/g.
\end{equation}
These beads can be interpreted as monopoles or antimonopoles trapped
on the string, and like kinks correspond to an element of $Z_2$.

When $m_1^2 = m_2^2$ and $\kappa> 2\lambda$, the monopoles split into
two ``semipoles'', whose possible fluxes, again inferred from the
differences between the fluxes of adjacent strings, are
\begin{equation}
\Phi_B = \{1,i,-1,-i \}4\pi/g. 
\end{equation}
Hence semipoles correspond to an element of $Z_4$.  Two adjacent
semipoles need not have total charge zero, and so need not annihilate.
An equivalent argument applies to the charge $w'$ in the case
$\kappa<2\lambda$.

Finally, when the $D_4$ symmetry is enlarged to a $\mathrm{O}(2)$ symmetry
(\ref{e:U1Sym}) at $\kappa = 2\lambda$ the monopoles become spread
out, and there is instead a continuous phase at each point of the
string, which is the argument of the complex parameter $w$ in
Eq.~(\ref{e:ComFluA}) or $w'$ in Eq.~(\ref{e:ComFluB}).
In general, the
phase can change along the string, giving rise to 
persistent currents, much like a superfluid.

These semipole and superfluid solutions are
new, and were not anticipated in Ref.\ \cite{Hindmarsh:1985xc}.

We have performed the first 3-dimensional numerical simulations of the
theory (\ref{e:SU2HLag}), in order to look for semipole solutions, and
to investigate the dynamics of networks of necklaces in the early
universe.  We take $\lambda=1/2$ and $g=1$ so that the string tension
is precisely $\pi v_2^2$ in the continuum.  We perform simulations on
$72^3$ periodic lattices with lattice spacing $a=1$, starting
  from random initial conditions. The
system is evolved with heavy damping until it has relaxed to straight
strings wrapping one of the directions in the simulation
volume. Details, and numerical simulations of the string network, will
be reported on in a separate publication~\cite{Hindmarsh:2016dha}.

Fig.~\ref{fig:isolated} shows isosurfaces of $\Tr \, \Phi_1^2$
(blue) and $\Tr \, \Phi_2^2$ (green) at $\frac{5}{16}v_{1,2}^2$, 90\%
isosurfaces of energy density (red), and streamlines of the vector
fields $B_i^{(1)}$ (blue) and $B_i^{(2)}$ (green).  In the general
case, shown at the top with $(a m_1)^2 = 0.1$, $(a m_2)^2=0.04$ and
$\kappa = 2\lambda$, one expects to see strings as tubes of constant
$\Tr \, \Phi_2^2$, with monopoles appearing as spheroids of constant $\Tr
\, \Phi_1^2$. It is clear that this expectation is borne out.

We also show the three different degenerate cases -- $\kappa =
4\lambda$, $2\lambda$, $\lambda$, with $(a m_1)^2 = (a m_2)^2 = 0.1$
-- from second to top downwards in Fig.~\ref{fig:isolated}. In the
$\kappa = 4\lambda$ case (second to top) the blue and green
streamlines are those of $B_i^{(+)}$ and $B_i^{(-)}$; otherwise they
show $B_i^{(1)}$ and $B_i^{(2)}$.

One sees that the semipoles are sources of the magnetic fields
$B_i^{(\pm)}$ ($\ka > 2\la$, second from top) or $B_i^{(1,2)}$
($\kappa < 2\lambda$, bottom), as explained above.

Finally, when $\kappa=2\lambda$, the accidental symmetry broken by the
string is $\mathrm{O}(2)$, and there is a 1-parameter family of solutions
labelled by the argument of $w$ (or equivalently $w'$).  Here, the energy density is uniform along the string.  

In this paper we have found a new class of topological objects on
strings in non-Abelian gauge theories, which we term semipoles.  They
are the result of the string breaking discrete symmetries of the
Lagrangian, and can be viewed as half of a
bead~\cite{Hindmarsh:1985xc}. In the $\mathrm{SU}(2)$ theory we study they can be
mapped to an element of $Z_4$.

The theory can be embedded in, for example, $\mathrm{SO}(10)$ with a
Higgs in the 126 \cite{Kibble:1982ae}, where symmetry enforces the
fields having the same mass scales and $\ka=2\la$, and so models the
string solution in an attractive class of Grand Unified
Theories~\cite{Jeannerot:2003qv,Allys:2015kge,Allys:2015yda}.  Note
that the complex flux we have used to classify the semipoles and
strings cannot be defined in every case.

It is plausible that accidental global symmetries will be
present in many GUT models, and that semipoles or superfluid degrees of freedom
are generic features
of high-scale cosmic strings.  
Such supercurrents might stabilise loops of string against collapse,
resulting in an exotic form of metastable matter
\cite{Copeland:1987th,Davis:1988ij}.

The cosmological implications of beads and semipoles depend on their
average separation along the string $d$.  It has been variously argued
that $d$ evolves to the string width \cite{Berezinsky:1997td} or
scales with the horizon \cite{BlancoPillado:2007zr}.  The latter
argument assumed that the objects living on the string would
annihilate if they encountered each other, which is not always the
case with semipoles. The presence of massive objects on the strings
affects the dynamics considerably \cite{Siemens:2000ty}.  The 
uncertainty in theory, observational signals and constraints motivates
the numerical investigation which we report on elsewhere
\cite{Hindmarsh:2016dha}.

\begin{acknowledgments}

Our simulations made use of the COSMOS Consortium supercomputer
(within the DiRAC Facility jointly funded by STFC and the Large
Facilities Capital Fund of BIS). DJW is supported by the People
Programme (Marie Sk{\l}odowska-Curie actions) of the European Union
Seventh Framework Programme (FP7/2007-2013) under grant agreement
number PIEF-GA-2013-629425. MH acknowledges support 
from the Science and Technology Facilities Council (grant
number ST/J000477/1).

We dedicate this paper to Tom Kibble.

\end{acknowledgments}

\bibliography{shortstrings}

\end{document}